\documentstyle[preprint,aps]{revtex}
\tighten
\begin{document}
\preprint{UAB-FT-362, gr-qc/9504049}
\draft

\title{Entanglement entropy in curved spacetimes
with event horizons}
\author{Rainer M\"uller\thanks{Electronic address:
rainer.mueller@physik.uni-muenchen.de}}
\address{Sektion Physik der Universit\"at M\"unchen,
Theresienstr. 37, D-80333 M\"unchen, Germany}
\author{and Carlos O. Lousto\thanks{Electronic address:
lousto@ifae.es~.
Permanent address: Instituto de
Astronom\'{\i}a y F\'{\i}sica del Espacio, Casilla de Correo 67 -
Sucursal 28, 1428  Buenos Aires, Argentina. Electronic address:
lousto@iafe.uba.ar}}
\address{IFAE--Grupo de F\'\i sica Te\'orica, Universidad Aut\'onoma
de Barcelona, E-08193 Bellaterra, Spain}
\date{\today}
\maketitle

\begin{abstract}

We consider the computation of the entanglement entropy
in curved backgrounds with event horizons. We use a Hamiltonian
approach to the problem and perform numerical computations on a
spherical lattice of spacing $a$. We study the cosmological case and
make explicit computations for the Friedmann-Robertson-Walker
universe.
Our results for a massless, minimally coupled scalar field
can be summarized by $S_{ent}=0.30 r_H^2/a^2$,
which resembles the flat
space formula, although here the horizon radius, $r_H$, is
time-dependent.

\end{abstract}


\section{Introduction}

The formal proof of the validity of the first law of black hole
thermodynamics\cite{SW92,IW94} for arbitrary perturbations of a
stationary
black hole in any gravitational theory
derivable from a diffeomorphism invariant
Lagrangean gave an impulse and renewed
interest on the computation of the black
hole entropy.
Recently, several authors have obtained
expressions for the {\it thermodynamic}
or coarse-grained
black hole entropy in the form of an
integral of the entropy density on any
cross-section of the Killing
horizon. More explicit expressions have been given for different
higher derivative gravitational (interacting with matter)
Lagrangeans\cite{EL94,CLA94,CLA95}. The main
techniques used to derive the black hole
entropy have been the Noether charge
construction\cite{W93}, the Euclidean
signature method\cite{V93}, and the field
redefinition approach\cite{JKM94}.

All the above derivations make use of the
first law of black hole thermodynamics where all its terms
are known (mass,  temperature, angular momentum, angular
velocity of the horizon, electromagnetic charge
and potential), but an integral on the horizon. One precisely
identifies this unknown quantity on the horizon with the
black hole entropy by analogy to any other ordinary
thermodynamic system. This can be called the coarse--grained
entropy because it is derived in the thermodynamic limit.
It is, however, desirable to have an
independent, statistical mechanical derivation of the entropy,
which may be called the fine--grained entropy. Relatively
early attempts\cite{tH85,ZT85} to compute this quantum-originated
entropy considered the {\it statistical} entropy,
$S\approx k_B\sum_np_n\ln p_n\approx k_B\ln N ,$
with N being the number of modes or
configurations of field very close to the black hole horizon.
't Hooft\cite{tH85} has computed the quantum corrections to
the black hole
entropy by counting the number of classical eigenmodes
of a scalar field
outside the event horizon. He immediately noted the appearance of an
ultraviolet divergence in the entropy as one approaches the  horizon
in counting the modes, and discussed the need of imposing a short
distance cut-off.

Bombelli {et al.} \cite{BKLS86}
studied instead the {\it entanglement} entropy
$S=-{\rm Tr}\left[\rho\ln\rho\right]$ by
considering in flat space a scalar quantum field in its ground
state and computing its reduced density
matrix, $\rho=|\Phi><\Phi|$, where one has
traced over the field degrees of
freedom inside a region of space that is intended to represent the
interior of a black hole.
Thus, the entanglement entropy is associated to correlations across
the boundary separating two regions.
More recently, several\cite{FN93,S93,CW94,KS94,D94,F94} computations
of the entanglement entropy agree in obtaining the
proportionality to the area $A$ of the
surface dividing the known from the unknown
regions and the $1/h^2$ ultraviolet divergence, where $h$ is the
minimal distance to the horizon we reach in counting
modes of the external field,
typically given in Minkowskian or Schwarzschild radial coordinates.
Their results can be summarized by
\begin{equation}
S=g{A\over \gamma h^2}~,
\label{1.1}
\end{equation}
where $g$ accounts for the number of fields and its helicities and
$\gamma=1/(360\pi)$ (see Refs. \cite{D94,F94}).

In Refs.\ \cite{FN93,F94,BFZ94} the entropy is computed by
identifying the
dynamical degrees of freedom with the states of quantum fields
propagating
in the black hole's interior. In Refs.\ \cite{tH85,D94} the
entropy of fields outside the black hole is studied instead.
The two kinds of
computation are conciliated in Ref.\ \cite{KS94} where it is shown
that
the density matrices of the two situations are identical.

The question of the ultraviolet divergence of the entropy is
discussed in
Ref.\ \cite{DLM95} where the Pauli-Villars regularization is used to
avoid
the introduction of a Planck scale cut-off
in the computation of the entropy. In the
paper\ \cite{HLW94} the entropy difference
between two states is shown
to be better behaved. Susskind\ \cite{S94b}
speculates about the possibility
that the black hole entropy be finite in string theory.

In the present paper, we handle this problem,
following Ref.\ \cite{S93}, by
making computations on a lattice of spacing $a$,
what renders the entropy finite.
In Srednicki's work the reduced ground state density
matrix of a free, massless, scalar quantum field is obtained by
tracing over the degrees
of freedom located inside of an imaginary
sphere of radius $R$ and by making
use of the analogy with the result\ \cite{BKLS86} for a coupled
system of
harmonic oscillators. The final result is
\begin{equation}
S={0.30\over4\pi}\left({4\pi R^2\over a^2}\right)
\label{1.2}
\end{equation}
which, as the author remarks, bears an striking similarity
with the formula for
the coarse--grained entropy of black holes
\begin{equation}
S={A_H\over 4 l_{\text{Pl}}^2},
\label{1.3}
\end{equation}
where $A_H$ is the area of the black hole's event horizon.
The similarity between these two equations is in fact surprising,
since
Srednicki's computation has been carried out in {\it flat} spacetime
while
we know that the geometry generated by a black hole is curved. The
argument
here is that the modes which contribute mostly to the entropy are
those
of short wavelengths that do not ``feel'' whether the spacetime is
curved
\ \cite{BKLS86}. However, the event horizon is a null surface which
greatly differs from the spacelike surface $R={\text{constant}}$. In
addition, we expect the entanglement entropy of a quantum field
computed in the curved
metric of a black hole to give {\it just} a correction (due to
quantum
fluctuations) to the coarse-grained entropy produced by the
classical
curved black hole background, but not the whole value (given by
Eq.\ (\ref{1.3}).)

Comparatively much less attention has been paid to the cosmological
case. Here one can also define a horizon temperature and the laws of
thermodynamics \cite{GH77,D87}, although their interpretation is not
as transparent as in the black hole case. In our paper we will
concentrate
on the entanglement correction to the classical entropy of the
universe.
In Sec. II we generalize the approach of Ref.\ \cite{S93}
to the case
when the spacetime is curved and possesses a horizon acting as a barrier
for our knowledge of the quantum state beyond it. In Sec. III we
review some properties of the Friedmann-Robertson-Walker geometry
while in
Sec. IV we give our main results. The paper finishes with a
discussion of the cosmological results and it possible extension
to the black hole case.

\section{Computation of entanglement entropy in curved spacetime}

In this paper, we want to extend the existing schemes for calculating
the entanglement entropy to a curved spacetime with horizon. To
compute the entanglement entropy, one considers a quantum field in
its
ground state. The spacetime is divided into two parts which are
separated by a boundary (in spacetimes with horizon,
this subdivision appears quite naturally). A given observer has
access
to the information of only one of these two regions.
In his/her subsystem, the field cannot be described
by a pure state wave function. Because of the correlations that
exists between the two regions, even in the ground state of the
field,
it must be described by a density matrix with an associated nonzero
entropy. This is the origin of the entanglement entropy.

In Refs \cite{BKLS86,S93} a scheme was given for calculating
the entanglement entropy in Minkowski space. They considered the
quantum field as a set of coupled harmonic oscillators with
Hamiltonian given by
\begin{equation} H = {1\over 2} \sum_i \Pi_i^2 +
	{1\over2}\sum_{i,j}\phi_i K_{ij} \phi_j .
	\label{eqSred7}\end{equation}
In the usual flat spacetime there is not horizon and the
boundary can be chosen arbitrarily. From (\ref{eqSred7}), it
is then possible to compute the entanglement entropy
numerically\cite{S93}.

We consider a curved spacetime with an isotropic metric of
the following form
\begin{equation} ds^2 = C^2(r,t)dt^2 - B^2(r,t)(dx^2 + dy^2 + dz^2) .
\label{eq19}\end{equation}
For a massive, non minimally coupled scalar field $\phi$, the
Lagrangean is given by
\begin{equation} {\cal L}= {1\over 2}\sqrt{-g} \left[ g^{\mu\nu}
(\partial_\mu\phi)(\partial_\nu \phi)
-(m^2+\xi R)\phi^2
\right] , \label{eq18}\end{equation}
which becomes for the metric (\ref{eq19})
	\begin{equation} {\cal L}= {1\over2} C B^3
	\left[ C^{-2}(\partial_t \phi)^2
	- B^{-2}(\nabla \phi)^2
-(m^2+\xi R)\phi^2
	\right]. \label{eq33}\end{equation}
The canonical momentum conjugate to $\phi$ is
\begin{equation} \Pi = {\partial
{\cal L}\over \partial (\partial_0 \phi)}
	=\sqrt{-g} g^{0\mu}\partial_{\mu} \phi
	= C^{-1} B^3 \partial_t \phi .\label{eq21}\end{equation}
We obtain the Hamiltonian of the scalar field from the definition
\begin{equation} H = \int d^3x
\left[ \Pi \partial_t \phi - {\cal L} \right],
	\label{eq22}\end{equation}
which becomes in our case
\begin{equation} H= {1\over 2}\int d^3x \left[ C B^{-3} \Pi^2 +
CB (\nabla \phi)^2
+CB^3(m^2+\xi R)\phi^2
\right] . \label{eq23}\end{equation}

The problem can be reduced to an effectively one-dimensional one
if we exploit its spherical symmetry. The metric
(\ref{eq19}) has an isotropic form then decomposition in
terms of spherical harmonics is possible. Following Srednicki
\cite{S93} in
using the real spherical harmonics $Z_{lm}(\theta,\varphi)$ which
are orthonormal and complete. We decompose the field and the
canonical momentum into partial wave components
\begin{eqnarray} \phi(\vec x) &=& {1\over r}\sum_{lm}
Z_{lm}(\theta,\varphi)\phi_{lm}(r), \label{eq24}\\
	\Pi (\vec x) &=& {1\over r}\sum_{lm} Z_{lm}(\theta,\varphi)
	\Pi_{lm}(r). \label{eq25}\end{eqnarray}
When we substitute (\ref{eq24}) and (\ref{eq25}) into (\ref{eq23}),
we obtain after some algebra
\begin{eqnarray}
H={1\over2} \sum_{lm} \int_0^\infty dr \Bigg[&&C B^{-3} \Pi_{lm}^2
+ C B \left({\partial \phi_{lm} \over \partial r}\right)^2\cr
&&+\left({\partial\over\partial r} C B \right){1\over r}
\phi_{lm}^2+C B {l(l+1)\over r^2} \phi_{lm}^2
+CB^3(m^2+\xi R)\phi_{lm}^2
\Bigg] .\label{eq26}\end{eqnarray}
We next perform a substitution in the canonical
variables in such a way that the commutation relations are
preserved:
\begin{equation}
\Pi'_{lm}= F\Pi_{lm}, \qquad \phi'_{lm}={1\over F}\phi_{lm},
\label{eq27}\end{equation}
with $F(r,t)=C^{1/2} B^{-3/2}$. In terms of the variables
$\phi'_{lm}$
and $\Pi'_{lm}$, our Hamiltonian takes the form of Eq. (\ref{eqSred7})
\begin{eqnarray}
H = {1\over 2} \sum_{lm} \int_0^\infty dr \Bigg[&&{\Pi'}^2_{lm}
+ C B \left({\partial \over \partial r} C^{1\over 2}B^{-{3\over 2}}
\phi'_{lm} \right)^2 \cr
&&+ {C\over B^{3}}
\left({\partial\over \partial r}
C B \right) {1\over r} {\phi'}^2_{lm} + {C^2\over B^{2}}
{l(l+1)\over r^2} {\phi'}^2_{lm}
+C^2(m^2+\xi R){\phi'}_{lm}^2
\Bigg] .\label{eq28}\end{eqnarray}
We can write this as
\begin{eqnarray}
H={1\over2} \sum_{lm}\int_0^\infty dr \Bigg[&&{\Pi'_{lm}}^2
	+{(\gamma_1+\gamma_4)\over a^2}{\phi'_{lm}}^2
	+{\gamma_2\over a}
	{\phi'_{lm}}{\partial\over\partial r}\phi'_{lm}\cr
	&&+ \gamma_3
	\left({\partial \phi'_{lm}\over \partial r}\right)^2
	+\gamma_5 {l(l+1)\over r^2} {\phi'_{lm}}^2
+\gamma_6 {\phi'}_{lm}^2
	\Bigg] ,\label{eq28a}\end{eqnarray}
where
$$\gamma_1(r)=a^2 CB \left({\partial\over\partial r}
	C^{1\over2} B^{-{3\over2}}\right)^2,
	\qquad \gamma_2(r)=2a C^{3\over2}B^{-{1\over2}}
{\partial\over
	\partial r}\left(C^{1\over2}B^{-{3\over2}}\right), $$
\begin{equation} \gamma_3(r)=\gamma_5(r)=C^2B^{-2}, \qquad
\gamma_4(r)={a^2\over r}CB^{-3}{\partial\over\partial r}(CB)
{},\qquad \gamma_6(r)=C^2 a^2(m^2+\xi R)
{}.\label{eq28b}\end{equation}
In Eqs (\ref{eq28a}) and (\ref{eq28b}), we have introduced an
arbitrary length scale $a$ to make the $\gamma_i$ dimensionless.
In the form (\ref{eq28a}), the Hamiltonian is in a suitable form for
discretization.

In order to compute the entanglement entropy
numerically, we put the system
on a spherical lattice with $N$ points in radial direction.
The lattice spacing is denoted $a$ so that $r=j a$, where $j$
is an integer. Discretization of (\ref{eq28a}) leads to
\begin{eqnarray}
H={1\over2a} \sum_{jlm} \Bigg[&&\Pi_{lmj}^2 + \left(\gamma_1+\gamma_4
	+\gamma_5{l(l+1)\over j^2}
+\gamma_6
	\right) \phi^2_{lmj}\cr
	&&+ \gamma_2 \phi_{lmj}
	(\phi_{lmj+1}-\phi_{lmj}) +
	\gamma_3(\phi_{lmj+1}-\phi_{lmj})^2
	\Bigg] .\label{eq28c}\end{eqnarray}
To discretize the parameters $\gamma_i$, it is necessary to specify
the form of the spacetime metric. This will be considered in the
next sections. The Hamiltonian (\ref{eq28c})
is of the general form (\ref{eqSred7}) with
$$ K_{j,j}=\gamma_1(j)-\gamma_2(j)+\gamma_3(j+1)+\gamma_3(j)
	+\gamma_4(j)+\gamma_5(j){l(l+1)\over j^2}+\gamma_6(j)
{},$$
\begin{equation}
K_{j,j+1}=K_{j+1,j}={1\over2}\gamma_2(j)-\gamma_3(j). \end{equation}

Knowing the explicit form of the $\gamma_i$, it is thus immediately
possible to calculate numerically the
entanglement entropy using Srednicki's
algorithm \cite{S93}: The lattice is divided into two parts.
We want to trace over the first $n$ lattice sites (the region inside
the horizon). To this end, the matrix $K$ is written via singular
value decomposition as $K=U^T K_D U$ where $K_D$ is diagonal and $U$
orthogonal. One defines the square root of $K$ as $\omega= U^T
K_D^{1 \over 2} U$ and writes it in the form
\begin{equation}
\omega = \left( \begin{array}{cc} A & B \\ B^T & C \end{array}
	\right) ,\end{equation}
where the submatrix $A$ is $n \times n$ and $C$ is $(N-n) \times
(N-n)$. Next we introduce $\tilde B={1 \over 2} B^T A^{-1} B$ and
$\gamma = C-\tilde B$. We perform a singular value decomposition
of $\gamma$: $\gamma=V^T \gamma_D V$. We finally define the matrix
$\tilde B'=\gamma_D^{-{1\over 2}} V \tilde B V^T
\gamma_D^{-{1\over 2}}$
which has $N-n$ eigenvalues $\tilde B_i'$. The entanglement entropy
associated with the trace over the $n$ sites inside the
horizon is
\begin{equation} S = \sum_{l=0}^\infty (2l+1) S_l ,\end{equation}
where
\begin{equation}
S_l = \sum_{i=1}^{N-n} \left[ \log (1-\xi_i) - {\xi_i \over 1-\xi_i}
\log \xi_i \right] ,\end{equation}
and $\xi_i = \tilde B'_i / (1+ \sqrt{1-{\tilde{B'}_i}^2})$. A
derivation
of this algorithm as well as its physical discussion can be found in
\cite{S93}.

\section{Geometry of Robertson-Walker universes}

It is our aim in this paper to calculate the entanglement entropy
of a quantum field in a curved spacetime with horizon.
Specifically, we are interested in the behavior of the entropy
for different classes of cosmological models.
We consider therefore the
Friedmann-Robertson-Walker (FRW) line element
\begin{equation} ds^2 = d t^2 - \beta^2 C^2( t)
\left[d\chi^2 +{\sin^2(\sqrt{k}\chi)\over k} (d\theta^2
+\sin^2\theta d\varphi^2)\right] ,\label{eq10}\end{equation}
and treat the three different cases $k=1, -1, 0$ corresponding to
a closed, open and spatially flat universe. The constant $\beta$
has the dimension of a length and determines the spacetime
curvature.

In the form (\ref{eq10}), the metric is
not very useful for calculating the entanglement entropy
because it is
not in the isotropic form (\ref{eq19}). It can obtained
by means of the coordinate transformation \cite{M73}
\begin{equation} d\eta = C^{-1}( t) d t, \qquad
x = {2 \beta \over\sqrt{k}} \tan \left(\sqrt{k}{ \chi\over 2}\right)
\sin\theta \cos\varphi, \label{eq11}\end{equation}
$$y = {2 \beta\over\sqrt{k}} \tan \left(\sqrt{k}
{\chi\over 2}\right) \sin\theta\sin\varphi, \qquad
z = {2 \beta\over\sqrt{k}} \tan \left(\sqrt{k}{ \chi\over 2}\right)
\cos\theta. $$
Note that it does not mix space and time variables.
For the sake of simplicity, we restrict ourselves in the following to
times $\eta>0$.

With (\ref{eq11}), the line element becomes
\begin{equation} ds^2 = C^2 ( \eta)\left[ d\eta^2 -
	{1\over \left(1+k { r^2\over4 \beta^2}\right)^2 }
	(d x^2 + d y^2 + d z^2)\right].\label{eq13}\end{equation}
It is of the form (\ref{eq19}) and it will be the starting point for
our subsequent calculation of the entanglement entropy.

An important concept for our discussion is the notion of the
cosmological event horizon. It represents the maximal
coordinate distance a photon emitted at time $t$ can travel.
For a metric of the form (\ref{eq10}), it is given by\cite{D88}
\begin{equation}
r_H = {2\beta\over\sqrt{k}}
\tan\left(\sqrt{k}{\Delta\eta\over2\beta}\right) ,
	\label{eq21a}\end{equation}
where
\begin{equation}
\Delta\eta = \eta(t_f)-\eta(t) =
\int_{t}^{t_f} {dt'\over {C(t')}}.
\label{eq7}\end{equation}
Here, $t_f$ equals infinity for $k=0,-1$, while for $k=+1$ it
corresponds to the time when the universe collapses to a
final singularity.

We will be interested in the area spanned by the cosmological event
horizon. Generally, in a curved spacetime with diagonal metric,
the proper area of a spherical surface $ r= r_0=\hbox{constant}$
is defined by \cite{W84}
\begin{equation}
A_0 = \int_{ r= r_0} \sqrt{g_{\theta\theta} g_{\varphi\varphi}}
	d\theta d\varphi. \label{eq5}\end{equation}
For the metric (\ref{eq13}), this becomes
\begin{equation}
A _0=4\pi C^2(t) {\sin^2( \sqrt{k}\chi_0)\over k}=4\pi C^2(\eta)
{ r_0^2\over \left(1+k{ r_0^2\over4 \beta^2}\right)^2 }.
\label{eq15}\end{equation}

As a concrete example, one can consider the evolution of a universe
with negligible matter contribution but endowed with a cosmological
constant $\Lambda$.  In that case, we have for $k=\pm 1, 0$ the
following evolution
\begin{equation}
C(t) = {1\over 2} \left( e^{t/\beta} + k e^{-t/\beta} \right) ,
\end{equation}
where $\beta = \sqrt{3/\Lambda}$. In terms of the conformal time
\begin{equation}
\Delta\eta={2\beta\over\sqrt{k}}\arctan
\left[\sqrt{k}e^{-t/\beta}\right] ,
\end{equation}
we have
\begin{equation}
C(\eta) = {\sqrt{k}\over 2} \left[ \tan\left(-\sqrt{k}
{\Delta\eta\over2\beta}\right) +
\tan^{-1}\left(-\sqrt{k}{\Delta\eta\over 2\beta}\right)\right] .
\end{equation}
The ``radius'' of the cosmological horizon can then be explicitly
computed from (\ref{eq21a}):
\begin{equation}
r_H = {2\beta \over k} \left( C-\sqrt{C^2 -k}\right).
\label{eq7a}\end{equation}
The area of the horizon can be calculated by inserting (\ref{eq7a})
into (\ref{eq15}). We obtain
\begin{equation} A_H = 4\pi \beta^2. \label{eq17}\end{equation}
It is noteworthy that it does not depend on $C(\eta)$, i. e. it is
time independent.

In the $k=0$ case, the evolution law is
$C(t)=\exp(t/\beta)/2$, corresponding
to $C(\eta)=\beta/\Delta\eta$. The horizon radius is
\begin{equation} r_H = \Delta\eta = {\beta \over C} ,
\label{eq8}\end{equation}
and the area is again given by (\ref{eq17}).

\section{Entanglement entropy in Friedmann-Robertson-Walker
universes}

With the formalism described in
Sec. II and the geometry reviewed
in Sec. III, it is now possible to
calculate numerically the entanglement
entropy of the quantum field which is
associated with the FRW cosmological
horizon. Let us first note that the
metric (\ref{eq13}) has the form (\ref{eq19}) with
\begin{equation} C = C(\eta), \qquad B= {C(\eta) \over
	1+k { r^2\over4 \beta^2} }. \label{eq32}\end{equation}
Evaluating Eqs. (\ref{eq28b}), we find that the coefficients
$\gamma_i$ are
$$ \gamma_1= {9k^2\over 16}{j^2\over b^4}, \qquad
\gamma_2={3k\over8}{j(kj^2+4b^2)\over b^4} ,$$
\begin{equation}
\gamma_3=\gamma_5=\left(1+k {j^2\over4 b^2}\right)^2, \qquad
\gamma_4=-{k\over 8}{kj^2+4b^2 \over b^4}
{},\qquad \gamma_6=C^2 a^2 \left[m^2-{6\xi\over C^3}
(\ddot C+kC)\right]
{}. \end{equation}
Here, the dimensionless quantity $b$ is defined by $b=\beta/a$.
Note that in the case $m=0=\xi$ the $\gamma_i$ are independent
of $\eta$, as will also be the Hamiltonian. We shall now
consider this simpler case.

In Minkowski space, the radius up to which the field has to
be traced over has been chosen arbitrarily
\cite{S93}. The entropy was then calculated for different
values of this radius. In a spacetime with horizon, however,
this procedure is not appropriate. The horizon sets a natural
radius for the region to be traced over since it represents a
boundary for the information an observer inside (in the cosmological
case, outside for a black hole) can have
about the outside region (inside for a black hole).
We therefore proceed in the
following way: At a fixed value of $C(\eta)$, we consider a set
of universes with different values of $\beta$. According to
(\ref{eq7a}),  they possess different horizon radii $r_H$.
We calculate the entanglement entropy numerically for each of
these universes by tracing up to the horizon radius (\ref{eq7a})
(this is equivalent to tracing for radius bigger than the horizon
due to the validity of the identity of the density matrices in
both cases \cite{KS94}). For a special value of the parameter $C$,
the result is plotted in Fig.\ \ref{figa} as a function of the
proper area of the
horizon which depends on $\beta$ via (\ref{eq17}). We find straight
lines in all of the cases $k=\pm 1, 0$.
This shows that the entanglement entropy
is proportional to the horizon area for the three spacetime classes:
\begin{equation}
S=  f_k(C) {A_H\over 4\pi a^2} .\label{eq32a}\end{equation}
The slope of these curves is given by the {\it a priori} unknown
quantity $f_k (C)/(4 \pi a^2)$. It depends on $k$ as well
as on the expansion parameter $C(\eta)$. We can try to find the
functional form of $f_k(C)$ by repeating the above procedure for
different values of $C$ and plotting $f_k$ versus $C$. Doing so,
one finds the results shown in Fig.\ \ref{figb} where the dots
represent our numerical computation of $f(C)$ for $k=\pm 1, 0$.

It would be desirable to know the analytical form of the functions
$f_k(C)$ and therefore of the entanglement entropy. Within our
numerical approach, we can only try to find a function that matches
our data points sufficiently well. In Fig.\ \ref{figb}, the solid
lines are plots of the functions
\begin{equation}
f_k(C)\cong 0.30 {r_H^2\over \beta^2} = 0.30
\cdot {4\over k^2}(C-\sqrt{C^2-k})^2.
\label{eq32b}\end{equation}
We see that the numerical values are perfectly fitted within
the numerical precision. Note that there are no free fitting
parameters apart from the numerical factor $0.30$. For
$k=0$  we recover the flat space Hamiltonian and
$f(C)\cong 0.30 C^{-2}$. The first equation in
(\ref{eq32b}) holds for general FRW spaces, while the second
equation, with the particular dependence on $C(\eta)$,
corresponds to the simplified case where we can neglect the
matter contribution to the evolution of the universe
(de Sitter--like geometry).

Taking into account (\ref{eq7a}) and (\ref{eq17}), this can be
written
\begin{equation} S_{ent}=0.30 {r_H^2\over a^2}.
\label{eq33b}\end{equation}
This simple formula is our final result for the open, asymptotically
open and closed FRW universe. Formally,
it resembles very much the Minkowski space formula (\ref{1.2}),
although one should keep in mind the conceptual differences
due to the curvature of the spacetime and the existence of an
event horizon, which are also reflected
in its specific calculation, as explained above.

\section{Discussion}

As we have seen, in the conformal time gauge for the metric, Eq
(\ref{eq13}), the Hamiltonian of a massless, minimally coupled
scalar field is time independent. However, since the boundary
($r_H$) up to which we trace over depends on time,
the entanglement entropy varies as the universe evolves.
In particular, it will increase as the universe
grows and decrease as it shrinks (for $k=1$) in the same way the
boundary of the observer's knowledge (given by the
cosmological event horizon) does. The dependence on
the particular way we take the slices of time has its origin in
the Hamiltonian approach we have used.


It is also interesting to recall here
that in Ref \cite{EL94} it was found,
for black holes in quadratic theories of
gravitation, also a proportionality between the entropy and the
square of the event horizon radius rather than to the horizon area.
Thus, in some sense, Eq (\ref{eq33b}) seems to be a more fundamental
relation than Eq (\ref{1.1}). Note however, that the above
proportionality for black holes in quadratic theories holds in the
conformal approximation to the solution made in Ref \cite{EL94}, but
it may well not survive in an exact solution (see Refs
\cite{CLA94,CLA95}). The other possibility to further investigate
this point is to directly apply the
procedure developed in this paper to the black hole case. This is
presently under study by the authors.

\begin{acknowledgments}
The authors thanks for kind hospitality the Fakult\"at f\"ur
Physik der Universit\"at
Konstanz where part of this work was carried out.
C.O.L was supported by the Direcci\'on General de
Investigaci\'on Cient\'\i fica y T\'ecnica of the Ministerio
de Educaci\'on y Ciencia de Espa\~na and CICYT AEN 93-0474,
and R. M. by the Deutsche Forschungsgemeinschaft.
\end{acknowledgments}

\begin{figure}
\caption{This figure shows the proportionality of our computed
entanglement entropy to the cosmological horizon area in a given
slice of time, i.e. for fixed conformal factor $C(\eta)$. Here
 $k=1, -1, 0$ correspond respectively to
a FRW closed, open and spatially flat universe.}
\label{figa}
\end{figure}

\begin{figure}
\caption{Here we plot the function $f_k(C)$ appearing in
Eq (\protect\ref{eq32a}) and defined by the proportionality
entropy--horizon area.
The dots correspond to our numerical computation while the
continuous line to the fit by $0.30 r_H^2/\beta^2$. The $k$
parameter reflects the effect of different spatial
curvatures. The dependence on $C(\eta)$ reflects time dependent
character of the de Sitter metric.}
\label{figb}
\end{figure}



\begin{references}

\bibitem{SW92} D.Sudarsky and R.M.Wald,
Phys. Rev. D {\bf  46}, 1454 (1992).\par

\bibitem{IW94} V.Iyer and R.M.Wald,
Phys. Rev. D {\bf  50}, 846 (1994).\par

\bibitem{EL94} A. Economou and C. O. Lousto, Phys. Rev. D,
{\bf 49}, 5278 (1994).\par

\bibitem{CLA94} M.Campanelli, C. O. Lousto and J. Audretsch,
Phys. Rev. D, {\bf 49}, 5188 (1994).\par

\bibitem{CLA95} M.Campanelli, C. O. Lousto and J. Audretsch,
Phys. Rev. D, {\bf 52},  (1995).\par

\bibitem{W93} R.M.Wald, Phys. Rev. D {\bf  48}, 3427 (1993).\par

\bibitem{V93} M.Visser, Phys. Rev. D {\bf  48}, 5697 (1993).\par

\bibitem{JKM94} T.Jacobson, G.Kang and R.C.Myers,
Phys. Rev. D {\bf  49}, 6587 (1994).\par

\bibitem{tH85} G.'t Hooft, Nucl. Phys. B, {\bf 256}, 727 (1985).\par

\bibitem{ZT85} W.H.Zurek and K.S.Thorne, Phys. Rev. Lett. {\bf 54},
2171 (1985).\par

\bibitem{BKLS86} L.Bombelli, R.K.Koul, J.Lee and R.D.Sorkin,
Phys. Rev. D, {\bf 34}, 373 (1986).\par

\bibitem{FN93} V.P.Frolov and I.D.Novikov, Phys. Rev. D,
{\bf 48}, 4545 (1993).\par

\bibitem{S93} M.Srednicki, Phys. Rev. Lett. {\bf 71}, 666 (1993).\par

\bibitem{CW94} C.Callan and F.Wilczek, Phys. Lett. B, {\bf 333},
55 (1994).\par

\bibitem{KS94} D.Kabat and M.J.Strassler, Phys. Lett. B, {\bf 329},
46 (1994).\par

\bibitem{D94} J.S.Dowker,  Class. Quantum Grav., {\bf 11}, L55
(1994).\par

\bibitem{F94} V.P.Frolov, preprint ALBERTA-THY-22-94,
gr-qc/9406037.\par

\bibitem{BFZ94} A.O.Barvinsky, V.P.Frolov and A.I.Zelnikov,
Phys. Rev. D, {\bf 51}, 1741(1995).\par

\bibitem{DLM95} J.-G.Demers, R.Lafrance and R.Myers, preprint
McGill/95-06, gr-qc/9503003.\par

\bibitem{HLW94} C.Holzhey, F.Larsen and F.Wilczek,
Nucl. Phys. B, {\bf 424}, 443 (1994).\par

\bibitem{S94b} L.Susskind, hep-th/9309145.\par

\bibitem{GH77} G.W.Gibbons and S.W.Hawking, Phys. Rev. D,
{\bf 17}, 2738 (1977).\par

\bibitem{D87} P.C.W.Davies, Class. Q. Gravity, {\bf 4}, L225
(1987).\par

\bibitem{M73} B.Mashhoom, Phys. Rev. D,
{\bf 8}, 4297 (1973).\par

\bibitem{D88} P.C.W.Davies, Ann. Inst. Henri Poincar\'e,
{\bf 49}, 297 (1988).\par

\bibitem{W84} R.M.Wald, {\it General relativity}, University
of Chicago Press, Chicago, 1984.\par


\end{references}
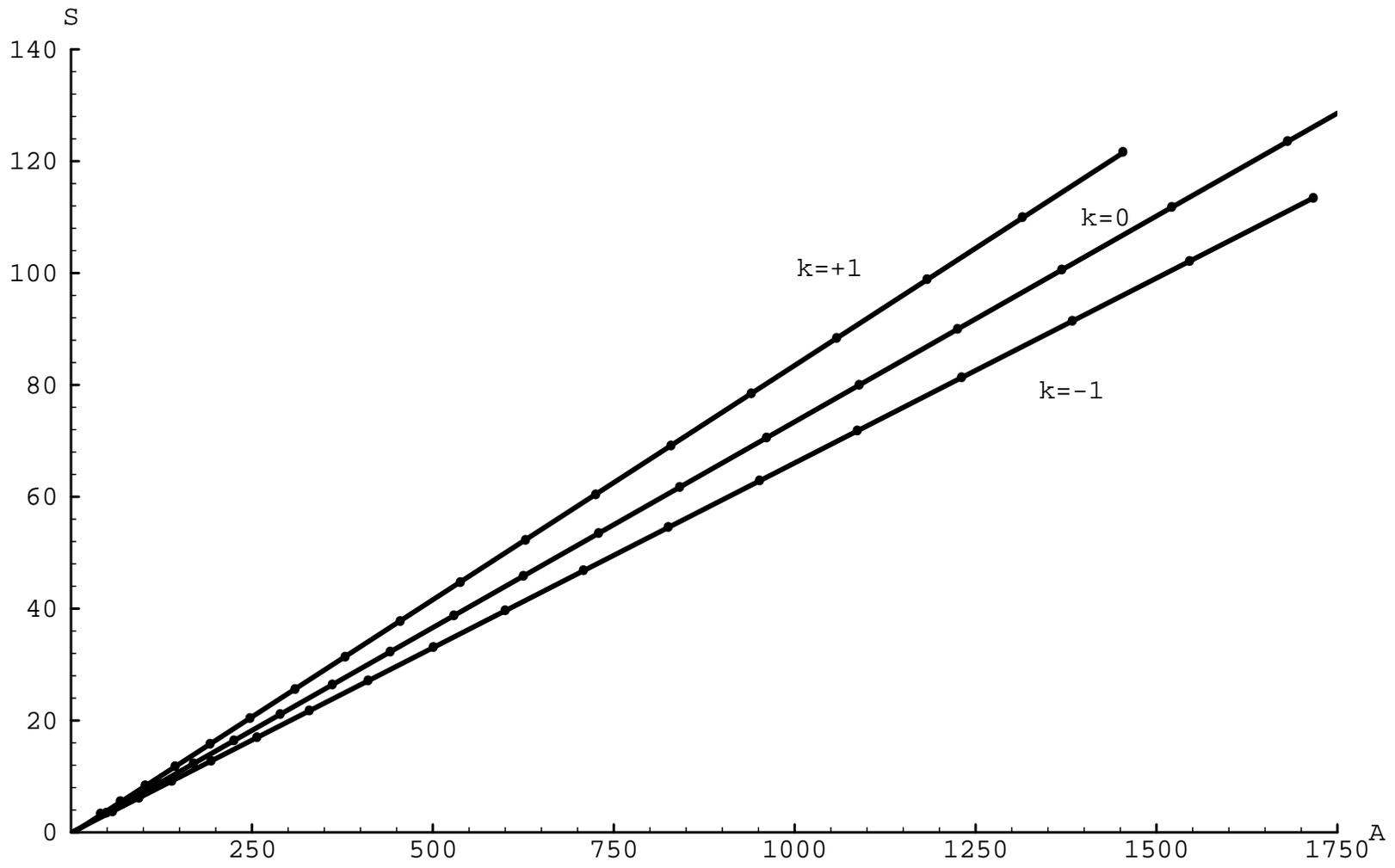
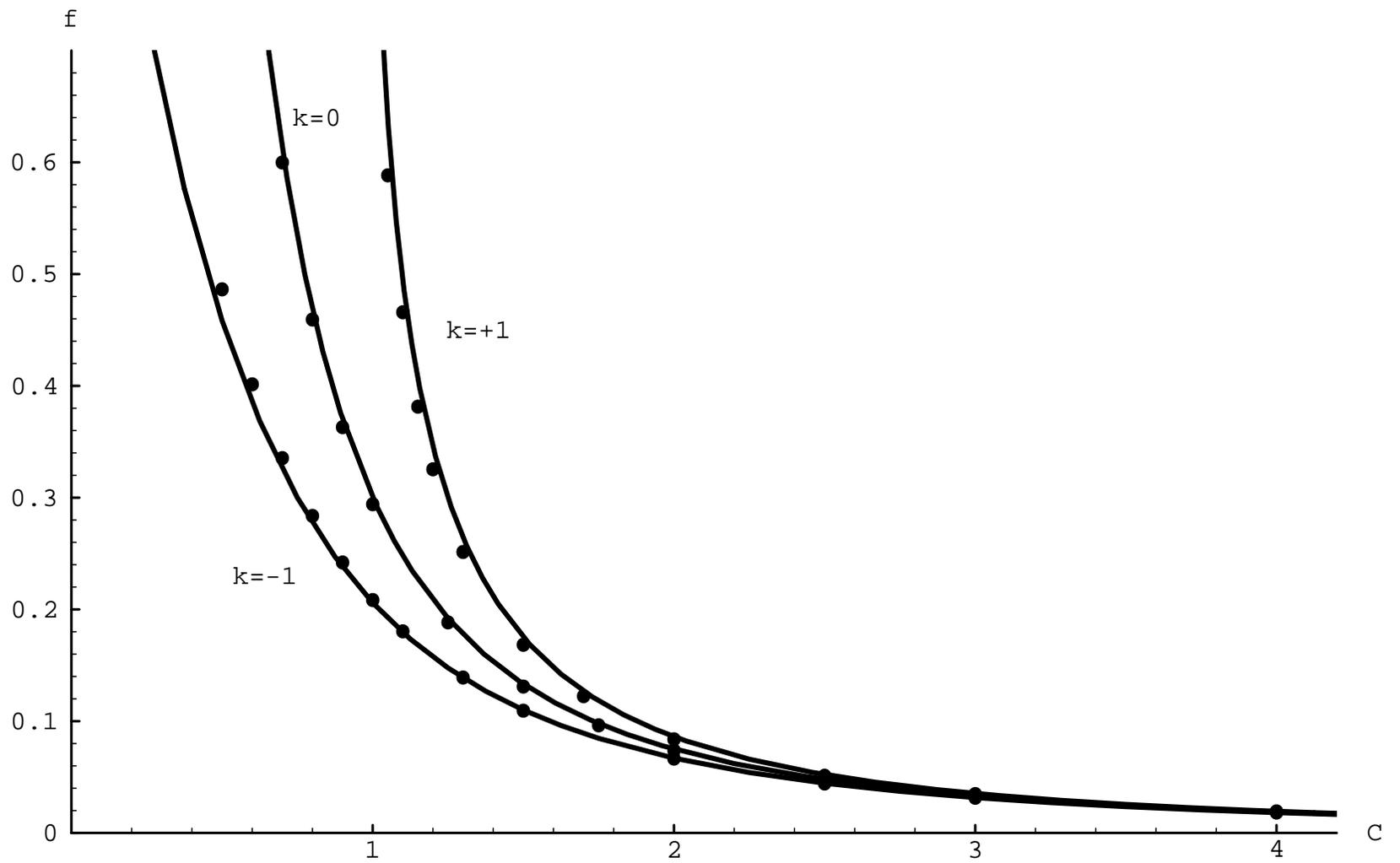
\end{document}